\documentclass[a4paper,12pt]{article}
\usepackage[T1]{fontenc}
\usepackage[utf8]{inputenc}
\usepackage[affil-it]{authblk}
\usepackage{slashed}
\usepackage{tikz}
\usetikzlibrary{decorations.text,calc,arrows.meta}
\usepackage{tkz-euclide}
\usepackage{caption}
\usetikzlibrary{perspective}
\usepackage{amsfonts, amssymb , amsmath, amsthm}
\usepackage{thm-restate}
\usepackage{nicefrac}
\usepackage{enumerate} 
\usepackage{enumitem}
\usepackage{xcolor}
\usepackage{a4wide}
\usepackage{bbding}
\usepackage{upgreek}
\usepackage{mathrsfs} 
\usepackage{bbm}
\usepackage{stmaryrd}
\usepackage{hyperref}
\hypersetup{
	colorlinks,%
	citecolor=green!50!black,%
	filecolor=black,%
	linkcolor=blue,%
	urlcolor=cyan!50!blue
}

\usepackage[numbers,sort&compress,merge]{natbib}
\usepackage{graphicx}

\usepackage{graphicx}
\usepackage[toc,page]{appendix}

\newcommand{\scalarProdTwo}[2]{ \left\langle {#1} | {#2} \right\rangle }

\newcommand{\A}{\mathbb{A}}

\newcommand{\ka}{\mathfrak{a}}

\newcommand{\kb}{\mathfrak{b}}

\newcommand{\sC}{\mathscr{C}}

\newcommand{\cD}{\mathcal{D}}
\newcommand{\D}{\mathbb{D}}

\newcommand{\kD}{\mathfrak{D}}
\newcommand{\rd}{\mathrm{d}}

\DeclareMathOperator{\re}{e}

\newcommand{\Fourier}{\mathcal{F}}

\newcommand{\R}{\mathbb{R}}
\newcommand{\C}{\mathbb{C}}

\DeclareMathOperator{\ri}{i}
\DeclareMathOperator{\tr}{tr}

\newcommand{\fN}{\mathsf{N}}

\newcommand{\cH}{\mathcal{H}}

\newcommand{\km}{\mathfrak{m}}
\newcommand{\N}{\mathbb{N}}

\newcommand{\cS}{\mathcal{S}}

\newcommand{\fU}{\mathsf{U}}

\newcommand{\Z}{\mathbb{Z}}

\newcommand{\Rd}{\R^{d}}

\newcommand{\bbS}{\mathbb{S}}

\newcommand{\symb}[1]{ \sigma_{\ms{#1}} }

\newcommand{\fP}{\mathsf{P}}
\newcommand{\fp}{\mathsf{p}}

\DeclareMathOperator{\Hess}{Hess}

\newcommand{\one}{\mathbbm{1}}

\newcommand{\coTanRd}{\mathrm{T}^{*} \mathbb{R}^{d}}

\newcommand{\xxi}{(x, \xi)}
\newcommand{\yeta}{(y, \eta)}

\DeclareMathOperator{\IHO}{\ms IHO}

\DeclareMathOperator{\Spec}{Spec}
\DeclareMathOperator{\ptSpec}{Spec_{pt}}

\newcommand{\ihcut}{\nicefrac{\ri}{\hbar} \,}

\DeclareMathOperator{\csch}{csch}
\DeclareMathOperator{\sech}{sech}

\newcommand{\ms}{\scriptscriptstyle}

\newcommand{\PsiDO}[1]{\Psi\mathrm{DO}^{{#1}} (\Rd)}

\title{Semi-classical Imprint of Horizon Induced Instability}
\author[1]{Arnab Chakraborty\thanks{arnab.chakraborty01@northsouth.edu}}
\author[2]{Onirban Islam\thanks{onirban.islam@math.uni-potsdam.de}}
\author[3]{Arshad Momen\thanks{arshad@iub.edu.bd}}
\affil[1]{Department of Mathematics and Physics, North South University, Dhaka-1229, Bangladesh}
\affil[2]{Institut f\"{u}r Mathematik, Universit\"{a}t Potsdam, 14476 Potsdam, Germany}
\affil[3]{Department of Physical Sciences, Independent University, Dhaka-1229, Bangladesh}
\date{17 May 2026}
\begin{document}
\maketitle
\begin{abstract}
We consider an inverted harmonic oscillator in the space $L^{2} (\bbS)$ of square-integrable functions on the circle $\bbS$ and compute its density of states employing the stationary phase approximation.  
Our computation is based on an oscillatory integral representation of the Schwartz kernel of the time-evolution operator.   
This demonstrates thermalisation as a semi-classical manifestation of the classical Lyapunov instability --- reported earlier 
in~\cite{Dalui_PRD_2020, Dalui_ii_PRD_2020} 
using heuristic analytic continuation. 
Our spectral analysis of the Hamiltonian points out and closes the conceptual and mathematical gaps in the preceding literature. 
\end{abstract}
%
%
%
%
%
%
%
%
%
%
\section{Introduction}


The Hawking 
temperature~\cite{Hawking_CMP_1975} 
is one of the quintessential predictions in semi-classical gravity. 
A few mechanisms (see the 
reviews~\cite{Wald_LivRevRel_2001, Hollands_PR_2015, Kay_EncyMathPhys_2025})  
have been offered so far in order to understand its physical underpinning. 
Several years ago,  
Dalui and his 
collaborators~\cite{Dalui_PRD_2020, Dalui_ii_PRD_2020} 
have shown that this temperature can be understood as the semi-classical manifestation of the Lyapunov instability of a classical particle around a black hole horizon. 
To be precise, they have argued that the classical Hamiltonian for a $1$-dimensional chargeless and massless particle in a neighbourhood of a certain class of (e.g., static spherically symmetric) black hole horizons is given by $h \xxi := \kappa x \xi$ in the leading-order approximation as long as only the radial motion is of interest. 
Here, $\kappa$ is the surface gravity of the horizon, $x$ is the radial distance of the particle from the horizon, and $\xi$ is the momentum of the particle. 
This is a Berry-Keating Hamiltonian which is related to the classical Hamiltonian $h_{-}$ of an inverted harmonic oscillator (IHO) via a canonical transformation, i.e, $h$ is symplectomorphic to $h_{-}$. 
The phase-space dynamics (given by the Hamiltonian flow) $\varPhi_{t}^{-}$ of an IHO is known to be instable in the sense of Lyapunov stability (see 
e.g.~\cite[Rem. 9.8.6]{Rudolph_Springer_2013},~\cite{Hashimoto_PRD_2017, Dalui_PRD_2020, Dalui_ii_PRD_2020}). 
This explains the instability of classical particles around a black hole horizon. 
Quantum mechanically, the Berry-Keating Hamiltonian is unitary equivalent to an IHO Hamiltonian $\hat{H}_{-}$. 
Dalui \textit{et al}.~\cite{Dalui_PRD_2020, Dalui_ii_PRD_2020} 
have computed the density of states $\uprho_{-}$ an IHO using the Gutzwiller trace 
formula~\cite{Gutzwiller_JMP_1971, *Gutzwiller_Springer_1990, *Gutzwiller_Scholarpedia_2007} 
(see 
also~\cite{Meinrenken_ReptMathPhys_1992} 
and the 
exposition~\cite{Uribe_Cuernavaca_1998}), 
and found that $\uprho_{-}$ exhibits the behaviour of a canonical ensemble at the Hawking temperature. 
Since only closed phase-space trajectories (Hamiltonian orbits) contribute to the Gutzwiller trace formula, their computation of $\uprho_{-}$ connects instability to thermalisation as the leading-order process.  

There are, however, a few conceptual and computational gaps in their work. 
As remarked  
in~\cite[after (6)]{Dalui_PRD_2020},  
an IHO does not admit any periodic phase-space trajectory (Hamiltonian orbit) $\gamma^{-} (t)$, and hence they have performed the analytic continuation $\upomega \mapsto \ri \upomega$ in the characteristic frequency $\upomega$ in order to apply the Gutzwiller trace formula to compute the semi-classical Green function of a simple harmonic oscillator (SHO). 
Under this rotation, an IHO Hamiltonian $h_{-}$ is \textit{formally} mapped to a SHO Hamiltonian $h_{+}$. 
In other words, Dalui and his 
collaborators~\cite{Dalui_PRD_2020, Dalui_ii_PRD_2020} 
have extracted the density of states $\uprho_{+}$ of a quantum SHO from its semi-classical Green function, and then obtained $\uprho_{-}$ by analytically continuing $\uprho_{+}$. 
The key step in their approach --- analytic continuation --- is unsatisfactory on both physical and mathematical grounds for the reasons explained below. 

Recall that, for each fixed energy $E (>0)$, an IHO phase-space trajectory $\gamma_{l}^{-} (t)$ comprises two $l = 1, 2$ distinct hyperbolic branches corresponding to exponentially diverging motions in opposite directions. 
In contrast, there is only one closed elliptical trajectory $\gamma^{+} (t)$ for a SHO. 
Therefore, $\uprho_{+} (E)$ is given by only summation over $\gamma^{+}$s according to the Gutzwiller trace formula whereas $\uprho_{-} (E)$ requires an additional summation over $l$. 
The latter is missing 
in~\cite[(15)]{Dalui_PRD_2020} 
(see 
also~\cite[(C5)]{Dalui_ii_PRD_2020}). 

As mentioned above, the primary motivation to perform the analytic continuation is to evade the lack of closed trajectories of an IHO. 
But this is equivalent to considering time as an imaginary parameter as can be seen from the following consideration.   
The phase-space dynamics $\varPhi_{t}^{+}$ of a SHO is periodic with the primitive period 
\begin{equation} \label{eq: period_SHO}
    T := \frac{2 \pi}{\upomega}  
\end{equation} 
whilst $\varPhi_{t}^{-}$ is not periodic at all.  
If the period is allowed to be imaginary then, of course, the time-evolution $\varPhi_{t}^{-}$ becomes periodic with primitive period $2 \pi \ri / \upomega$ which is precisely the Wick-rotated version of $T$. 
Observe that this is equivalent to analytic continuation in time $t \mapsto \ri t$. 
But then the trajectories $\gamma_{l}^{-} (\ri t)$ of an IHO are \textit{not} instable anymore. 
Therefore, the analytic continuation, $\R \times \R$ as the phase-space, and time as an imaginary parameter are not consistent when instability and periodicity are required simultaneously. 
This, combined with the energy degeneracy of $\gamma_{l}^{-}$, in particular, shows that the Maslov index $\km_{l}^{-}$ supposed to appear in the semi-classical expression of $\uprho_{-} (E)$ is not that $\km^{+}$ appearing in the semi-classical expression of $\uprho_{+} (E)$, and hence the Maslov index 
in~\cite[(15)]{Dalui_PRD_2020} 
(see 
also~\cite[(C5)]{Dalui_ii_PRD_2020}) 
is incorrect. 

The analytic continuation furthermore yields completely misleading spectral properties of an IHO.  
Recall that the spectrum of a quantum SHO Hamiltonian $\hat{H}_{+}$ consists of discrete positive energy eigenvalues $E_{n}^{+}, n \in \N \cup \{ 0 \}$ with square-integrable eigenfunctions (given by the Hermite polynomials). 
Upon the Wick rotation, this yields \textit{complex} eigenvalues $\lambda_{\nu}, \nu \in \C$ associated with \textit{non}-square-integrable eigenfunctions (given by parabolic cylindrical functions) for an IHO. 
Therefore, the operator one formally obtains from the analytic continuation of $\hat{H}_{+}$ \textit{cannot} be interpreted as the Hamiltonian observable of an IHO, and hence it is unclear at this stage whether the quantity Dalui and his collaborators~\cite[(15)]{Dalui_PRD_2020} 
(see 
also~\cite[(C5)]{Dalui_ii_PRD_2020}) 
have obtained as the analytic continuation of $\uprho_{+} (E)$ actually corresponds to some observable (density of states) of an IHO. 

The purpose of this article is two-fold: to resolve the above-mentioned conceptual and mathematical gaps, i.e, provide the appropriate set-up for an IHO so that its density of states $\uprho_{-}$ makes sense physically, and compute $\uprho_{-}$ without using the analytic continuation. 
To circumvent the conflict between the instability and the periodicity of an IHO orbits $\gamma_{l}^{-} (t)$, we consider time $t$ as a \textit{complex} parameter $\tau$ and extend the relevant quantities holomorphically on the strip 
\begin{equation} \label{eq: strip_holo}
    \D := \{ \tau = t + \ri s \in \C \,|\, s \in (0, \pm \beta) \}. 
\end{equation}
It now turns out that the phase-space dynamics $\varPhi_{\tau}^{-}$ is periodic in its imaginary part as explained in Section~\ref{sec: class_instability}. 
Note that this set-up is also required in order to define a thermal (quantum) state (see 
e.g.~\cite{Fulling_PR_1987, Bratteli_Springer_1997}), 
and hence it is natural owing to the subject matter of this article.  

The spectral theory of an IHO requires an intricate treatment. 
The non-square-integrability of the eigenfunctions of an IHO implies that we do not have a $L^{2} (\R)$-Hilbert space to begin with, rather one must consider the quantisation $H_{-}$ of $h_{-}$ as an operator on tempered distributions $\cS' (\R)$ because parabolic cylindrical functions belong to this space. 
It turns out that $H_{-}$ is \textit{formally} self-adjoint (see~\eqref{eq: def_formal_adjoint_op}), and hence the existence of complex eigenvalues $\lambda_{\nu}$ is not a contradiction with the spectral theory of self-adjoint 
operators\footnote{Note 
    that $\cS' (\R)$ is not a Hilbert space and hence the notion of self-adjoint operators (see~\eqref{eq: def_symm_op}) does not make sense.}
on $L^{2} (\R)$. 
As a consequence, $H_{-}$ cannot be considered as the Hamiltonian observable of an IHO. 
We invoke the Gel'fand triplet formalism (see Appendix~\ref{sec: Gelfand_triplet}) to define the Hamiltonian observable $\hat{H}_{-} : L^{2} (\R) \to L^{2} (\R)$, and analyse its spectrum in Section~\ref{sec: spec_th}. 
A standard result (see, 
e.g.~\cite[Satz 24.6]{Weidmann_Springer_2003}) 
from spectral theory entails that $\hat{H}_{-}$ has \textit{purely absolutely continuous} real spectrum which shows that the density of states for $\hat{H}_{-}$ does not make sense albeit we have the correct Hamiltonian observable $\hat{H}_{-}$. 
We therefore consider an IHO on the circle $\bbS$ in order to ensure discrete spectrum of $\hat{H}_{-} : L^{2} (\bbS) \to L^{2} (\bbS)$. 
Notice that, this means $\bbS \times \R$ is our classical phase space, and then closed Hamiltonian orbits $\gamma_{l}^{-} (\tau)$ are guaranteed. 

We compute the density of states $\uprho_{-}$ of an IHO in the semi-classical regime employing the stationary phase approximation originally due to Wentzel, Kramers, Brillouin, and Jeffreys (see 
e.g.~\cite[Prop. 5.2]{Dimassi_CUP_1999},~\cite[Thm. 3.11]{Zworski_AMS_2012},~\cite[Thm. 12.8.4]{Rudolph_Springer_2013}). 
Since the self-adjointness of the Hamiltonians $\hat{H}_{\pm}$ is at our disposal, we have the time-evolution operators $U_{t}^{\pm} := \re^{- \ihcut t \hat{H}_{\pm}}$ by the spectral theorem. 
Then the density of states $\uprho_{\pm}$ of a(n) S(I)HO can be computed from the Fourier transform 
\begin{equation} \label{eq: DoS_Fourier_tr_time_evolution_op_SIHO} 
    \uprho_{\pm} (E) = \Fourier_{t \mapsto E}^{-1} (\tr_{\cH_{\pm}} U_{t}^{\pm}), \quad \cH_{+} := L^{2} (\R), \quad \cH_{-} := L^{2} (\bbS)    
\end{equation}
of the (vacuum) partition functions $\tr_{\cH_{\pm}} U_{t}^{\pm}$. 
The caveat of this formal computation is that $U_{t}^{\pm}$ are \textit{not} trace-class. 
As a consequence, 
all\footnote{In 
    general, there can be infinitely many singularities.
}
the singularities in $\tr_{\cH_{\pm}} U_{t}^{\pm}$ might be contributing to the Fourier transform. 
To circumvent this problem, one must view the partition functions as  distributions and characterise their singularity structure. 
The precise characterisation of this class of distributions was carried out by 
Duistermaat and Guillemin~\cite{Duistermaat_InventMath_1975} 
in great 
generality\footnote{The 
    authors considered the operator $U_{t} := \re^{- \ri t P}$ on a closed manifold, where $P$ is a first-order positive elliptic pseudodifferential operator, and proved that $U_{t}$ is a Fourier integral 
    operator~\cite[Thm. 1.1]{Duistermaat_InventMath_1975}. 
} 
when $\hbar = 1$. 
Their work was subsequently propounded to the semi-classical setting by
Meinrenken~\cite{Meinrenken_ReptMathPhys_1992}. 
Their analysis requires to express the Schwartz kernels $\fU_{t}^{\pm}$ of $U_{t}^{\pm}$ as oscillatory integrals. 
This is well documented for a SHO and is known as the Mehler kernel. 
The Mehler kernel $\fU_{t}^{-}$ of an IHO is usually presented from the analytic continuation viewpoint in physics 
literature~\cite[Sec. 4.1]{Barton_AnnPhys_1986}, 
and it is hard to find a concrete reference for its systematic derivation albeit the Mehler kernel for a generic quadratic form is reported in mathematics  
literature~\cite{Derezinski_JEDP_1993, Hoermander_MathZ_1995}. 
Therefore, we have presented a derivation of the Mehler kernels $\fU_{t}^{\pm}$ for both simple and inverted harmonic oscillators based on the semi-classical Weyl quantisation (see Appendix~\ref{sec: PsiDO}) in Section~\ref{sec: Mehler_kernel} since the derivation proceeds in parallel for both cases. 
This also paves the way to characterise the singularities of $\fU_{t}^{\pm}$ --- an essential requirement in order to use~\eqref{eq: DoS_Fourier_tr_time_evolution_op_SIHO}. 

The advantage of computing $\uprho_{-} (E), E >0$ employing the stationary phase lemma instead of the Gutzwiller trace formula is its simplicity and concreteness albeit both yields the same result. 
Our computaton in Section~\ref{sec: DoS_IHO} exhibits that Maslov factor $\km_{l}^{-}$ and the number of Hamiltonian orbits $\gamma_{l}^{-} (\tau)$ appearing in the semi-classical expression (see~\eqref{eq: DoS_IHO_WKB}) of $\uprho_{-} (E)$ differs from that reported by Dalui and his 
collaborators~\cite[(15)]{Dalui_PRD_2020} 
(see 
also~\cite[(C5)]{Dalui_ii_PRD_2020}) 
for precisly the reasons elucidated above. 

%
%
%
%
%
%
%
%
%
%
\section{Simple and inverted harmonic oscillators}
\label{sec: SIHO}
\subsection{Classical dynamics and (in)stability} 
\label{sec: class_instability}

Recall that the classical Hamiltonian of a $1$-dimensional simple (resp. inverted) harmonic oscillator of characteristic frequency $\upomega$ is a real function 
\begin{equation} \label{eq: def_class_hamiltonian_SIHO}
	h_{\pm} \xxi := \frac{\xi^{2}}{2m} \pm \frac{m}{2} \upomega^{2} x^{2} 
\end{equation} 
on the phase space $\R \times \R$, where $x = x (t)$ and $\xi = \xi (t)$ are the position and momentum of a classical particle of mass $m$ at time $t$, respectively. 
Let 
\begin{equation} \label{eq: cst_energy_surface_SIHO}
    \Sigma_{E}^{+} := \{ \xxi \in \R \times \R \,|\, h_{+} \xxi = E \}, 
    \quad 
    \Sigma_{E}^{\IHO} := \{ \xxi \in \R \times \R \,|\, h_{-} \xxi = E \}
\end{equation} 
be the constant energy surfaces of these Hamiltonians. 
It is straightforward to see that the classical dynamics is governed by the canonical transformations ($h_{\pm}$-hamiltonian flows)  
\begin{align} \label{eq: Hamilton_flow_SIHO}
    \varPhi_{t}^{+} 
    = \left(
	\begin{array}{cc}
	      \cos \upomega t & \dfrac{\sin \upomega t}{m \upomega} 
	      \\ 
	      - m \upomega \sin \upomega t & \cos \upomega t
	\end{array}
	  \right), 
    \quad 
    \varPhi_{t}^{-} = \left(
    \begin{array}{cc}
	      \cosh \upomega t & \dfrac{\sinh \upomega t}{m \upomega} 
	      \\ 
	      - m \upomega \sinh \upomega t & \cosh \upomega t
	  \end{array}
	  \right). 
\end{align}
Locally these maps are parametrised by the generating functions (of the first kind)  
\begin{align} \label{eq: generating_function_Hamiltonian_flow_SIHO}
    S_{t}^{+} (x, y) 
    & := 
    \frac{m \upomega}{2} \big( (x^{2} + y^{2}) \cot \upomega t - 2xy \csc \upomega t \big), 
    \nonumber \\
    S_{t}^{-} (x, y) 
    & := 
    \frac{m \upomega}{2} \big( (x^{2} + y^{2}) \coth \upomega t - 2xy \csch \upomega t \big). 
\end{align}
The level sets of the Hamiltonians $h_{\pm}$, the orbits $\gamma^{\pm} (t)$ of $\varPhi_{t}^{\pm}$, and the constant energy surfaces $\Sigma_{E}^{+}$ and $\Sigma_{E}^{\IHO}$ coincide for both cases. 
Moreover, $\xxi = (0, 0)$ is the only critical point of the Hamiltonian vector fields $X_{\pm} = \xi / m \partial_{x} \mp m \upomega^{2} x \partial_{\xi}$. 

In the case of a harmonic oscillator, $\Sigma_{E > 0}^{+}$ is an ellipse whose origin is the element in $\Sigma_{0}^{+}$.  
The phase-space dynamics $\varPhi_{t}^{+}$ is periodic with period $kT$ where $T$ is the primitive period~\eqref{eq: period_SHO} and $k \in \Z$. 
Let $\gamma^{+} (t)$ be the phase-space trajectory (Hamiltonian orbit) of energy $E > 0$. 
Then the periodicity of $\varPhi_{t}^{+}$ is equivalent to 
\begin{equation} \label{eq: const_energy_periodic_geodesic_SHO}
    \forall k \in \Z: \quad h_{+} \big( \gamma^{+} (t + kT) \big) = E >0. 
\end{equation}
Since $\Sigma_{E}^{+}$ is compact and (simply-)connected for any (non-negative) energy $E$, the action along $\gamma^{+}$ yields (see 
e.g.~\cite[Def. 11.4.1, Thm. 11.4.7, and Exm. 11.5.1]{Rudolph_Springer_2013}
\begin{equation} \label{eq: action_angle_SHO}
    I^{+} = \int_{\Sigma_{E}^{+}} \xi \rd x = 2 \pi \frac{E}{\upomega}. 
\end{equation}

In contrast to a harmonic oscillator, the hamiltonian $h_{-}$ of an inverted harmonic oscillator is not necessarily non-negative, the constant energy surfaces $\Sigma_{E}^{\IHO}$ are hyperbolas whose origin is the element in $\Sigma_{0}^{\IHO}$, and its canonical transformation $\varPhi_{t}^{-}$ is not periodic. 
As mentioned in the Introduction, one of the pivotal motivations of this investigation is to put forward the thermodynamic interpretation (by 
Dalui et al~\cite{Dalui_PRD_2020} 
and 
Dalui and Majhi~\cite{Dalui_ii_PRD_2020}) 
of the density of states for an IHO in a rigorous setting. 
Therefore, let us briefly recall the notion of a thermal quantum state in the sense of the Kubo-Martin-Schwinger (KMS) condition. 
This condition means that the thermal (Wightman) $2$-point correlation functions have holomorphic continuations to the strip~\eqref{eq: strip_holo} such that they satisfies the KMS condition at $s = 0$ and $s = \beta$, where $\beta$ is a positive constant that physically represents the inverse temperature of the system. 
We refer to the 
expositions~\cite{Fulling_PR_1987, Bratteli_Springer_1997} 
and original references therein for details. 

We now consider the holomorphic extension (denoted by the same symbol) of the canonical transformation $\varPhi_{\tau}^{-}$ to $\D$ motivated by the setting of a thermal state. 
This means that the hyperbolic functions in~\eqref{eq: Hamilton_flow_SIHO} are simply replaced by their 
holomorphic extensions $\sinh \upomega \tau$ and $\cosh \upomega \tau$. 
A straightforward computation using the identities $\cosh (t + \ri s) = \cosh t \cos s + \ri \sinh t \sin s$ and $\sinh (t + \ri s) = \sinh t \cos s + \ri \cosh t \sin s$ show that the phase-space evolution is now periodic in its imaginary part of the flow parameter: 
\begin{equation} \label{eq: periodic_orbit_IHO}
    \varPhi_{\tau + \ri k T}^{-} = \varPhi_{\tau}^{-}. 
\end{equation}
Notice that $h_{-}$ is not non-negative and for each non-zero energy $E$, there are two distinct Hamiltonian orbits $\gamma_{l}^{-} (\tau), l = 1, 2$, i.e., 
\begin{equation} \label{eq: const_energy_periodic_geodesic_IHO}
    \forall k \in \Z, \quad l = 1, 2: \quad h_{-} \big( \gamma_{l}^{-} (\tau + \ri kT) \big) = E \gtrless 0. 
\end{equation}

%
%
%
%
%
%
%
%
%
%
\subsection{Spectral theory} 
\label{sec: spec_th}

We consider the differential 
operators\footnote{See Appendix~\ref{sec: Gelfand_triplet} for functional analytical concepts used in this section.}
\begin{equation} \label{eq: def_op_SIHO}
    H_{\pm} := - \frac{\hbar^{2}}{2m} \frac{\rd^{2}}{\rd x^{2}} \pm \frac{m}{2} \upomega^{2} \hat{x}^{2} : \cS (\R) \to \cS (\R)
\end{equation}
on the Schwartz space $\cS (\R)$ on $\R$, where $\hbar$ is the reduced Planck constant and $\hat{x}^{2}$ is the multiplication operator $x \mapsto x^{2}$. 
Note that these operators are continuous on $\cS (\R)$ which is dense in the space $\cH_{+}$ of square integrable (complex-valued) functions on $\R$. 

It is well-known that $H_{+}$ is essentially self-adjoint on $\cH_{+}$, and its unique self-adjoint extension $\hat{H}_{+}$ is the Hamiltonian of a quantum SHO whose spectrum~\eqref{eq: EV_SHO} consists of discrete energy eigenvalues 
\begin{equation} \label{eq: EV_SHO}
    \Spec \hat{H}_{+} = \{ E_{n}^{+} \,|\, n \in \N \cap \{ 0 \} \}, \quad E_{n}^{+} := \left( n + \frac{1}{2} \right) \hbar \upomega,    
\end{equation}
where $\hbar$ is the reduced Planck's constant. 

The operator $H_{-}$ is formally self-adjoint on $\cS (\R)$. 
Hence it acts on tempered distributions $\cS' (\R)$ on $\R$ by duality. 
To find the eigenvalues of $H_{-}$, one introduces the complex variable $z := \re^{\ri \pi / 4} \sqrt{2 m \upomega / \hbar} x$ so that the eigenvalue equation $H_{-} \psi = \lambda_{\nu} \psi$ yields the Weber differential equation   
\begin{equation} \label{eq: Weber_ODE}
    \frac{\rd^{2} \psi}{\rd z^{2}} + \left( - \frac{z^{2}}{4} - \ri \frac{E_{-}}{\hbar \upomega} \right) \psi (z) = 0. 
\end{equation}
The solution of this equation is well-known and is given by $\psi (z) = c_{1} \psi^{(1)} (z) + c_{2} \psi^{(2)} (z)$ where $\psi^{(1)}, \psi^{(2)}$ are parabolic cylindrical functions and $c_{1}, c_{2} \in \C$ are arbitrary constants. 
This yields complex eigenvalues 
\begin{equation} \label{eq: EV_IHO}
    \lambda_{\nu} := \left( \nu + \frac{1}{2} \right) \ri \hbar \upomega, \quad \nu \in \C, 
\end{equation}
which is precisely what one obtains from the analytic continuation of $E_{n}^{+}$. 
Observe that $\psi \notin \cH_{+}$ rather $\psi \in \cS' (\R)$. 
In other words, 
\begin{equation}
    H_{-} : \cS' (\R) \to \cS' (\R) 
\end{equation}
is \textit{not} the quantum Hamiltonian observable of an IHO since it is only formally self-adjoint, and hence its (complex) eigenvalues $\lambda_{\nu}$ cannot be interpreted as energy.  

In order to obtain the desired quantum Hamiltonian, one notes that the continuous inclusions $\cS (\R) \subset \cH_{+} \subset \cS' (\R)$ is a Gelfand triplet.  
The restriction $H_{- \upharpoonright \cH}$ of $H_{-}$ to $\cH$ is essentially self-adjoint 
by~\cite[Satz 17.15]{Weidmann_Springer_2003},  
and the fact that the space of compactly supported (complex-valued) functions on $\R$ is dense in $\cS (\R)$. 
Then the unique extension $\hat{H}_{-}$ of $H_{- \upharpoonright \cH}$ is the quantum Hamiltonian observable of an IHO. 
This operator has a purely absolutely continuous real spectrum (see, 
e.g.~\cite[Satz 24.6]{Weidmann_Springer_2003}) 
\begin{equation} \label{eq: spec_IHO}
    \Spec \hat{H}_{-} \subseteq \R. 
\end{equation}

As already mentioned in the Introduction, the primary purpose of this article is to investigate the density of states of an IHO. 
This requires discrete spectrum of $\hat{H}_{-}$. 
But we have just observed in~\eqref{eq: spec_IHO} that this requirement cannot be satisfied for an IHO on $\cH_{+}$. 
Therefore, we consider IHO on the phase space $\bbS \times \R$ and subsequently the quantum Hamiltonian as an operator 
\begin{equation}
    \hat{H}_{-} : \cH_{-} \to \cH_{-}
\end{equation}
on $\bbS$ in order to guarantee a discrete spectrum. 
Notice that this means 
\begin{equation} \label{eq: cst_energy_surface_IHO}
    \Sigma_{E}^{-} := \{ \xxi \in \bbS \times \R \,|\, h_{-} \xxi = E \}
\end{equation}
is the constant energy surfaces of an IHO pertinent to this article instead of $\Sigma_{E}^{\IHO}$ (\eqref{eq: cst_energy_surface_SIHO}).   

%
%
%
%
%
%
%
%
%
%
\subsection{Mehler kernels} 
\label{sec: Mehler_kernel}

The singularity structure of the time-evolution operators $U_{t}^{\pm}$ is best elucidated by their Schwartz kernels $\fU_{t}^{\pm}$ instead of the operators themselves. 
Therefore, the description of $U_{t}^{\pm}$ via the spectral theorem is inadequate for our purpose, and we require an oscillatory 
integral\footnote{These 
    integrals must be understood as distributions. 
    In fact, it is 
    well-known~\cite{Duistermaat_InventMath_1975, Meinrenken_ReptMathPhys_1992} 
    that these objections are Lagrangian distributions but we will not go into that direction to keep the technicalities bare minimum.
    } 
representation of $\fU_{t}^{\pm}$. 
This description of $\fU_{t}^{+}$ is well-documented for the simple harmonic oscillator and is known as the Mehler kernel. 
But a concrete reference for an inverted harmonic oscillator $\fU_{t}^{-}$ is not so easy to locate albeit the Mehler kernel for a generic quadratic form is 
reported~\cite{Derezinski_JEDP_1993, Hoermander_MathZ_1995}. 
We, therefore, spell out a derivation for both since the arguments proceed in parallel. 

The derivation is based on the Weyl calculus (see Appendix~\ref{sec: PsiDO}), i.e., our aim is to construct $U_{t}^{\pm}$ as the Weyl quantisation (see~\eqref{eq: Weyl_quantisation}) of $\varPhi_{t}^{\pm}$. 
Following, for 
instance~\cite[Sec. 1.7]{Taylor_AMS_1986},~\cite[Sec. 4]{Hoermander_MathZ_1995}, 
the strategy is to perform a Wick rotation in $t$ and solve the corresponding heat equation. 
In other words, we consider the solution operator $\tilde{U}_{t}^{\pm}$ of the initial-value heat equation  
\begin{equation} \label{eq: Schroedinger_heat_eq}
    \hbar \frac{\partial \tilde{U}_{t}^{\pm}}{\partial t} = - H_{\pm} \tilde{U}_{t}^{\pm}, 
    \quad 
    \tilde{U}_{0}^{\pm} = \one
\end{equation}
for non-negative $t$ and a fixed positive parameter $\hbar$. 
The Schwartz kernels of these operators are of the form (see~\eqref{eq: def_kernel_PsiDO})   
\begin{equation}
	\tilde{\fU}_{t}^{\pm} (\hbar; x, y) = \int_{\R} a_{t}^{\pm} \left( \frac{x+y}{2}, \xi \right) \re^{\ihcut (x-y) \xi} \frac{\rd \xi}{2 \pi \hbar}, 
\end{equation}
where $a_{t}^{\pm} \in \cS (\R \times \R)$ are called the Weyl symbols of $\tilde{U}_{t}^{\pm}$. 
Note that the Weyl calculus is covariant under the metaplectic group (see 
e.g.~\cite[Prop. 7.5, p. 74]{Taylor_AMS_1986} 
and that identity is known as the metaplectic covariance of
the Weyl calculus. 
Due to this symmetry it follows that $a_{t}^{\pm} = a_{t}^{\pm} (h_{\pm})$ is a function of $h_{\pm}$ only. 
Set $B_{t}^{\pm} := H_{\pm} \tilde{U}_{t}^{\pm}$ whose Weyl symbol is denoted by $b_{t}^{\pm}$. 
Since $h_{\pm}$ and $a_{t}^{\pm}$ Poisson commute, the product formula for Weyl symbols results  
\begin{equation}
    b_{t}^{\pm} \xxi 
    = 
    h_{\pm} \xxi a_{t}^{\pm} \big( h_{\pm} \xxi \big)  
    - \frac{\hbar^{2}}{8} \left( \frac{\partial^{2}}{\partial y \partial \xi} - \frac{\partial^{2}}{\partial x \partial \eta} \right)^{2} h_{\pm} \xxi \, a_{t}^{\pm} \big( h_{\pm} \yeta \big) \Big|_{y = x, \eta = \xi}.  
\end{equation}
A straightforward computation employing the chain rule and using the classical Hamiltonians~\eqref{eq: def_class_hamiltonian_SIHO} yields 
\begin{align}
    \left( \frac{\partial^{2}}{\partial y \partial \xi} - \frac{\partial^{2}}{\partial x \partial \eta} \right)^{2} h_{\pm} \xxi \, a_{t}^{\pm} \big( h_{\pm} \yeta \big) \Big|_{y = x, \eta = \xi} 
    & = 
    \frac{1}{m} \frac{\partial^{2} a_{t}^{\pm}}{\partial x^{2}} 
    \pm  
    m \upomega^{2} \frac{\partial^{2} a_{t}^{\pm}}{\partial \xi^{2}} 
    \nonumber \\ 
    & = 
    \pm 2 \upomega^{2} \left( h_{\pm} \frac{\partial^{2} a_{t}^{\pm}}{\partial h_{\pm}^{2}} + \frac{\partial a_{t}^{\pm}}{\partial h_{\pm}} \right)  
\end{align}
for the rightmost term of the preceding equation. 
Therefore~\eqref{eq: Schroedinger_heat_eq} entails  
\begin{equation}
    \hbar \frac{\partial a_{t}^{\pm}}{\partial t} 
    = 
    \left( \pm \frac{\hbar^{2} \upomega^{2}}{4} \frac{\partial^{2} a_{t}^{\pm}}{\partial h_{\pm}^{2}} - a_{t}^{\pm} \right) h_{\pm} 
    \pm \frac{\hbar^{2} \upomega^{2}}{4} \frac{\partial a_{t}^{\pm}}{\partial h_{\pm}}, 
    \qquad 
    a_{0}^{\pm} = 1. 
\end{equation}   
Motivated by the metaplectic covariance of the Weyl quantisation, we make the ansatz $a_{t}^{\pm} = f_{\pm} \re^{ \frac{2 g_{\pm}}{\hbar \upomega} h_{\pm}}$ where $f_{\pm}$ and $g_{\pm}$ are functions of $\upomega t/2$ and $f_{\pm}$ do not vanish identically. 
Then the initial condition enforces $f_{\pm} (0) = 1$ and $g_{\pm} (0) = 0$. 
After performing some straightforward computation, equating coefficients of $h_{\pm}$ results $g'_{\pm} = \pm g_{\pm}^{2} - 1$ and $f'_{\pm} / f_{\pm} = g_{\pm}$, where $g'_{\pm}$ and $f'_{\pm}$ denote derivatives of these functions with respect to their arguments $\upomega t/2$. 
Elementary hyperbolic and trigonometric identities entail $g_{+} = - \tanh (\upomega t/2), g_{-} = - \tan (\upomega t/2)$ and $f_{+} = \mathrm{sech} (\upomega t/2), f_{-} = \cos (\upomega t/2)$. 

The preceding construction holds for all $\Re t \geq 0$ and $\cosh (\upomega t/2) \neq 0$. 
Setting $t = \ri s$ with $s \in \R$ and renaming the latter by the former, one obtains the Schwartz kernels 
\begin{align} \label{eq: time_evolution_kernel_SIHO}
    & \fU_{t}^{\pm} = \ka_{\pm} (t) \int_{\R} \re^{\ihcut \varphi_{t}^{\pm}} \frac{\rd \xi}{2 \pi \hbar}, 
    \\  
    & \ka_{+} (t) := \frac{1}{\cos \frac{\upomega t}{2}}, 
    & \varphi_{t}^{+} (x, y, \xi) := (x-y) \xi - 2 \frac{h_{+}}{\upomega} \tan \frac{\upomega t}{2}, 
    \nonumber \\ 
    & \ka_{-} (t) := \frac{1}{\cosh \frac{\upomega t}{2}}, 
    & \varphi_{t}^{-} (x, y, \xi) := (x-y) \xi - 2 \frac{h_{-}}{\upomega} \tanh \frac{\upomega t}{2} 
    \nonumber
\end{align}
of $U_{t}^{+}, t \in \R \setminus \A$ and $U_{t}^{-}, t \in \R$, respectively, where $\A := \{ (n + 1/2) T \,|\, n \in \Z \}$ is the set of zeros of $\ka^{+} (t)$. 
Note that these integrals are oscillatory integrals, i.e., thy must be understood as distributions; see 
e.g.~\cite[Sec. 3.6]{Zworski_AMS_2012}.  
In particular, $\fU_{t}$ is a Schwartz distribution on $\R \times \R$: 
\begin{equation}
    \fU_{t}^{\pm} (\psi, \chi) = \frac{1}{2 \pi \hbar} \int_{\R \times \R \times \R} \ka_{\pm} (t) \re^{\ihcut \varphi_{t}^{\pm} (x, y, \xi)} \psi (x) \chi (y) \rd x \rd y \rd \xi 
\end{equation}
for $\psi, \chi \in \cS (\R)$. 

Viewing the oscillatory integrals~\eqref{eq: time_evolution_kernel_SIHO} as distribution, we want to characterise the singularity of these distributions. 
To proceed, the fibre-critical sets of $\varphi_{\pm}$ are given by (see 
e.g.~\cite{Duistermaat_InventMath_1975, Meinrenken_ReptMathPhys_1992},~\cite[App of Ch. 11]{Dimassi_CUP_1999}) 
\begin{equation} \label{eq: fibre_crit_mfd_SIHO}
    \sC_{t}^{+} = \Big\{ x = y + \frac{2 \xi}{m \upomega} \tan \frac{\upomega t}{2} \Big\}, 
    \quad 
    \sC_{t}^{-} = \Big\{ x = y + \frac{2 \xi}{m \upomega} \tanh \frac{\upomega t}{2} \Big\},   
\end{equation}
which yield the canonical relations 
\begin{equation} \label{eq: canonical_rel_SIHO}
    C_{t}^{\pm} := \{ (x, \rd_{x} \varphi_{t}^{\pm}; y, - \rd_{y} \varphi_{t}^{\pm}) \,|\, (x, y, \xi) \in \sC_{t}^{\pm} \}. 
\end{equation}
These  
sets\footnote{Geometrically 
    they are submanifolds. 
    In particular, $C_{t}^{\pm}$ are Lagrangian submanifolds; 
    see~\cite{Duistermaat_InventMath_1975, Meinrenken_ReptMathPhys_1992} 
    for details.}
(manifolds) encode the singularities of $U_{t}^{\pm}$ in the follow sense. 
The complement of $\sC_{t}^{\pm}$ is the set where $U_{t}^{\pm}$ are smooth. 
But $\sC_{t}^{\pm}$ does not provide any information along which directions (covectors) $U_{t}^{\pm}$ diverges. 
This is what $C_{t}^{\pm}$ captures: $U_{t}^{\pm}$ are singular at $x$ and $y$ in the directions $\rd_{x} \varphi_{t}^{\pm}$ and $- \rd_{y} \varphi_{t}^{\pm}$. 
In the language of microlocal analysis, $C_{t}^{\pm}$ are the twisted wavefronts of $U_{t}^{\pm}$ whereas $\sC_{t}^{\pm}$ are the singular supports of $U_{t}^{\pm}$.  

Recall that $\Fourier_{\xi \mapsto x-y}^{-1} \re^{\pm \ri a \xi^{2} / 2} = \sqrt{2 \pi a}^{-1} \re^{\mp \ri \xi^{2} / (2a)}$ for any non-zero real number $a$. 
Our convention of the (semi-classical) Fourier transform $\Fourier : \cS (\R) \to \cS (\R)$ is 
\begin{equation*}
	u \mapsto \tilde{u} := \Fourier (u) := \int_{\R} \re^{- \ihcut \, x \xi} u (x) \, \rd x, 
	\quad 
    \tilde{u} \mapsto u = \frac{1}{2 \pi \hbar} \int_{\R} \re^{\ihcut \, x \xi} \tilde{u} (\xi) \rd \xi.  
\end{equation*}
Applying this identity to~\eqref{eq: time_evolution_kernel_SIHO} in the $\xi$-variable yields the Mehler kernels 
\begin{align} \label{eq: Mehler_kernel_SIHO}
    \fU_{t}^{+} (x, y) 
    = 
    \kb_{\pm} (t) \re^{\ihcut S_{t}^{\pm} (x, y)}, 
    \quad  
    \kb_{+} (t) 
    = 
    \sqrt{\frac{m \upomega}{2 \pi \hbar \sin (\upomega t)}}, 
    \quad 
    \kb_{-} (t) 
    = 
    \sqrt{\frac{m \upomega}{2 \pi \hbar \sinh (\upomega t)}},  
\end{align}
where $S_{t}^{\pm}$ are defined in~\eqref{eq: generating_function_Hamiltonian_flow_SIHO}. 

\section{Density of states} 
\label{sec: DoS}

Since the spectrum~\eqref{eq: EV_SHO} of a simple harmonic oscillator on $\cH_{+} := L^{2} (\R)$ comprises discrete positive eigenvalues $E_{n}^{+}$, it makes sense to count the number $\fN_{+} (E) := \# \{ n \in \N \cup \{ 0 \} ~|~ E_{n}^{+} \leq E \}$ of eigenvalues up to an energy $E$. 
In contrast, the spectrum~\eqref{eq: spec_IHO} of an inverted harmonic oscillator on $L^{2} (\R)$ is \textit{continuous}.  
Therefore, we 
consider\footnote{This 
is obviously not the only way to ensure discrete eigenvalues. 
} 
the operator $\hat{H}_{-}$ on $\cH_{-} := L^{2} (\bbS)$ so that $\Spec \hat{H}_{-}$ is consists of \textit{discrete} real eigenvalues $E_{n}^{-}$. 
The Weyl counting function $\fN_{-} (E) := \# \{ n \in \N \cup \{ 0 \} ~|~ E_{n}^{-} \leq E \}, E >0$ for an inverted quantum harmonic oscillator then makes sense, and so does the density of states $\uprho_{-} (E) := \rd \fN_{-} / \rd E$. 

As mentioned in the Introduction, the density of states $\uprho_{\pm} (E)$ are the Fourier transformations~\eqref{eq: DoS_Fourier_tr_time_evolution_op_SIHO} of the vacuum parition functions $\tr_{\cH_{\pm}} U_{t}^{\pm}$. 
Note that these partition functions must be understood in the \textit{distributional} sense because $U_{t}^{\pm}$ are non-trace-class, as can be seen, for instance, from the fact that $U_{0}^{\pm}$ are the identity operators which are not clearly trace-class. 
In other words, the partition functions are viewed as continuous linear maps  
\begin{subequations} \label{eq: tr_U_t_dist}
    \begin{align}
        & \tr_{\cH_{+}} U_{t}^{+} : \cS (\R) \to \C, \quad \chi_{+} \mapsto \tr_{\cH_{+}} U_{\chi_{+}}^{+} := \tr_{\cH_{+}} \left( \int_{\R} U_{t}^{+} (\Fourier^{-1} \chi_{+}) (t) \rd t  \right), 
        \\ 
        & \tr_{\cH_{-}} U_{\tau}^{-} : \cS (\R) \to \C, \quad \chi_{-} \mapsto \tr_{\cH_{-}} U_{\chi_{-}}^{-} := \tr_{\cH_{-}} \left( \int_{\D} U_{\tau}^{-} (\Fourier^{-1} \chi_{-}) (\tau) \rd \tau \right), 
    \end{align}
\end{subequations}
where $\D$ is the strip given by~\eqref{eq: strip_holo} and the operators $U_{\chi_{\pm}}^{\pm}$ are given by 
\begin{subequations} 
    \begin{align}
        & U_{\chi_{}+}^{+} \psi := \left( \int_{\R} U_{t}^{+} (\Fourier^{-1} \chi_{+}) (t) \rd t \right) \psi, 
        \\ 
        & U_{\chi_{-}}^{-} \psi := \left( \int_{\D} U_{\tau}^{-} (\Fourier^{-1} \chi_{-}) (\tau) \rd \tau \right) \psi. 
    \end{align}
\end{subequations}
We remark that $(\Fourier^{-1} \chi_{-}) (\tau)$ is the holomorphic extension of inverse Fourier transform $(\Fourier^{-1} \chi_{-}) (t)$, i.e., 
\begin{subequations} 
    \begin{align}
        & (\Fourier^{-1} \chi_{+}) (t) = \frac{1}{2 \pi \hbar} \int_{\R} \re^{\ihcut t E} \chi_{+} (E) \rd E, 
        \\ 
        & (\Fourier^{-1} \chi_{-}) (\tau) = \frac{1}{2 \pi \hbar} \int_{\R} \re^{\ihcut \tau E} \chi_{-} (E) \rd E.  
    \end{align}
\end{subequations}

Since $\fU_{t} (x, y)$ is smooth in $x, y$ according to~\eqref{eq: fibre_crit_mfd_SIHO}, one can take the restriction $\fU_{t} (x, y)_{\upharpoonright x = y}$ to the diagonal $\{ x = y \}$ in order to compute the traces $\tr_{\cH_{\pm}} U_{t}^{\pm} = \int_{\R, \bbS} \fU_{t}^{\pm} (x, y)_{\upharpoonright x = y} \rd x$. 
Using this identity, the distributional formulation of~\eqref{eq: DoS_Fourier_tr_time_evolution_op_SIHO} yields  
\begin{subequations} \label{eq: DoS_Mehler_kernel_regularised}
    \begin{align}
        \int_{\R_{+}} \uprho_{+} (E) \chi_{+} (E) \rd E 
        & = 
        \tr_{\cH_{+}} U_{\chi_{+}}^{-}  
        = 
        \frac{1}{2 \pi \hbar} \int \ka_{+} (t) \re^{\ihcut \phi_{t}^{+} (x, \xi) + \ihcut t E} \chi_{+} (E) \, \rd t \, \rd x \, \rd \xi \, \rd E,   
        \\ 
        \int_{\R_{+}} \uprho_{-} (E) \chi_{-} (E) \rd E 
        & = 
        \tr_{\cH_{-}} U_{\chi_{-}}^{-}  
        = 
        \frac{1}{2 \pi \hbar} \int \ka_{-} (\tau) \re^{\ihcut \phi_{\tau}^{-} (x, \xi) + \ihcut \tau E} \chi_{-} (E) \, \rd \tau \, \rd x \, \rd \xi \, \rd E,   
    \end{align}
\end{subequations}
where $\phi_{t}^{\pm} (x, \xi) := \varphi_{t}^{\pm} (x, x, \xi)$. 
Since we are interested in the semi-classical limit of $\uprho_{\pm} (E)$, it is sufficient to evaluate the integral using the stationary phase approximation due to Wentzel, Kramers, Brillouin, and Jeffreys (see 
e.g.~\cite[Prop. 5.2]{Dimassi_CUP_1999},~\cite[Thm. 3.11]{Zworski_AMS_2012},~\cite[Thm. 12.8.4]{Rudolph_Springer_2013}). 

%
%
%
%
%
%
%
%
%
%
\subsection{Simple harmonic oscillator} 
\label{sec: DoS_SHO}

To begin with, we look for the critical points $(t_{0}, x_{0}, \xi^{0})$ of the phase function $\phi_{t}^{+} + tE$ of the oscillatory integral~\eqref{eq: DoS_Mehler_kernel_regularised}:  
\begin{subequations} \label{eq: stationary_condi_phase_func_SHO}
    \begin{align}
        \frac{\partial \phi_{t}^{+}}{\partial t} (t_{0}, x_{0}, \xi^{0}) 
        & = 
        - E 
        \Rightarrow 
        h_{+} (x_{0}, \xi^{0}) \sec^{2} \frac{\upomega t_{0}}{2} 
        = 
        E,  
        \label{eq: deri_phase_func_t_DoS_SHO}
        \\ 
        \frac{\partial \phi_{t}^{+}}{\partial x} (t_{0}, x_{0}, \xi^{0}) 
        & = 
        0 
        \Rightarrow 
        \frac{\partial \varphi_{t}^{+}}{\partial x} (t_{0}, x_{0}, x_{0}, \xi^{0}) + \frac{\partial \varphi_{t}^{+}}{\partial y} (t_{0}, x_{0}, x_{0}, \eta^{0})  
        = 
        \xi^{0} - \eta^{0} 
        = 0, 
        \label{eq: deri_phase_func_x_DoS_SHO} 
        \\ 
        \frac{\partial \phi_{t}^{+}}{\partial \xi} (t_{0}, x_{0}, \xi^{0}) 
        & = 
        0 
        \Rightarrow 
        \frac{\partial h_{+}}{\partial \xi} (x_{0}, \xi^{0}) \tan \frac{\upomega t_{0}}{2} 
        = 0, 
        \label{eq: deri_phase_func_xi_DoS_SHO} 
    \end{align}
\end{subequations}
where we have used~\eqref{eq: canonical_rel_SIHO} in~\eqref{eq: deri_phase_func_x_DoS_SHO}. 
Since $E$ is independent of $t$,~\eqref{eq: deri_phase_func_t_DoS_SHO} enforces $\sec^{2} (\upomega t_{0} / 2) = 1$ which entails 
\begin{equation}
    t_{0} = kT, \quad k \in \Z. 
\end{equation}
The stationary condition~\eqref{eq: deri_phase_func_x_DoS_SHO} shows that the initial trajectory $(y_{0}, \eta^{0})$ of the Hamiltonian orbit $\gamma^{+} (t)$ coincides with its final trajectory $(x_{0}, \xi^{0})$, i.e., $\gamma^{+} (t)$ is periodic. 
The last stationary condition~\eqref{eq: deri_phase_func_xi_DoS_SHO} does not impose any new constraint because $\tan (\upomega t_{0} / 2)$ vanishes identically at $t_{0}$. 
Therefore, $\phi_{t}^{+}$ satisfies the Hamilton-Jacobi equation at $\gamma^{+} (kT) \in \Sigma_{E}^{+}$; see~\eqref{eq: const_energy_periodic_geodesic_SHO} and~\eqref{eq: cst_energy_surface_SIHO}. 

Let $\Hess_{\gamma^{+} (t)} (\phi_{t}^{+} + tE)$ be the hessian of the phase function at $\gamma^{+} (t)$.  
Then $\Hess_{\gamma^{+} (t)} (\phi_{t}^{+} + tE) = \Hess_{\gamma^{+} (t)} \phi_{t}^{+}$ and a straightforward computation yields $\det \Hess_{\gamma^{+} (kT)} \phi_{t}^{+} = 2 \partial_{x} h_{+}$ $\big( \gamma^{+} (kT) \big) \partial_{\xi} h_{+} \big( \gamma^{+} (kT) \big) \neq 0$. 
That is, $\phi_{t}^{+} + tE$ is a non-degenerate phase function and the leading term in stationary phase approximation yields 
\begin{equation} \label{eq: DoS_SHO_WKB}
    \uprho_{+} (E) 
    \propto  
    \frac{1}{2 \pi \hbar \upomega} \sum_{\gamma^{+} (kT)} \re^{\ihcut kTE} \re^{\ri \pi \km^{+} / 4}, 
\end{equation}
where the constant of proportionality is a dimensionless constant and the signature of $\Hess_{\gamma^{+} (kT)} (\phi_{t}^{+} + tE)$ is known as the Maslov index $\km^{+}$ of $\gamma^{+}$.  
The Maslov index is well-known for a simple harmonic oscillator $\km^{+} = 2$ (see 
e.g.~\cite[Exm. 12.6.13 (1)]{Rudolph_Springer_2013}). 
It is a topological invariant and is equal to the Conley-Zehnder index. 
We refer to the 
textbook~\cite[Def. 7.7.1]{Rudolph_Springer_2013} 
and original references therein for details. 

We remark that the preceding expression is equivalent to the expression of $\uprho_{+}$ obtained via the Gutzwiller trace formula in the sense that they differ by some (dimensionless) constant. 
This can be seen by rewriting the non-Maslov contribution in~\eqref{eq: DoS_SHO_WKB} as $\re^{\ihcut k I^{+}}$ where $I^{+}$ is the action~\eqref{eq: action_angle_SHO} along $\gamma^{+}$. 


%
%
%
%
%
%
%
%
%
%
\subsection{Inverted harmonic oscillator} 
\label{sec: DoS_IHO}

The computation of $\uprho_{-}$ is structurally analogous to that described in Section~\ref{sec: DoS_SHO}. 
But there is a pivotal difference in achieving the critical point $t_{0}$. 
So we repeat some of the preceding steps for clarity. 
As before, the critical points $(t_{0}, x_{0}, \xi^{0})$ of the phase function $\phi_{t}^{-} + tE$ of the oscillatory integral~\eqref{eq: DoS_Mehler_kernel_regularised} are determined by  
\begin{subequations} \label{eq: stationary_condi_phase_func_DoS_IHO}
    \begin{align}
        \frac{\partial \phi_{t}^{-}}{\partial t} (t_{0}, x_{0}, \xi^{0}) 
        & = 
        - E 
        \Rightarrow 
        h_{-} (x_{0}, \xi^{0}) \sech^{2} \frac{\upomega t_{0}}{2}
        = 
        E, 
        \label{eq: deri_phase_func_t_DoS_IHO}
        \\ 
        \frac{\partial \phi_{t}^{-}}{\partial x} (t_{0}, x_{0}, \xi^{0}) 
        & = 
        0 
        \Rightarrow 
        \frac{\partial \varphi_{t}^{-}}{\partial x} (t_{0}, x_{0}, x_{0}, \xi^{0}) + \frac{\partial \varphi_{t}^{-}}{\partial y} (t_{0}, x_{0}, x_{0}, \eta^{0})  
        = 
        \xi^{0} - \eta^{0} 
        = 0,  
        \label{eq: deri_phase_func_x_DoS_IHO}
        \\ 
        \frac{\partial \phi_{t}^{-}}{\partial \xi} (t_{0}, x_{0}, \xi^{0}) 
        & = 
        0 
        \Rightarrow 
        \frac{\partial h_{-}}{\partial \xi} (x_{0}, \xi^{0}) \tanh \frac{\upomega t_{0}}{2} 
        = 0,  
        \label{eq: deri_phase_func_xi_DoS_IHO} 
    \end{align}
\end{subequations}
where we have used~\eqref{eq: canonical_rel_SIHO} in~\eqref{eq: deri_phase_func_x_DoS_IHO}. 
Since $E$ is independent of $t$,~\eqref{eq: deri_phase_func_t_DoS_IHO} enforces $\sech^{2} (\upomega t_{0} / 2) = 1$ which entails $t_{0} = 0$, i.e., we have only the \textit{trivial real} critical point in $t$. 
This is expected because the phase-space dynamics $\varPhi_{t}^{-}$ of an IHO is not periodic at all in contrast to that $\varPhi_{t}^{+}$ of a SHO. 

We circumvent this obstacle by extending the relevant quantities in $t \in \R$ holomorphically to $\tau$ in the strip~\eqref{eq: strip_holo}, motivated by the notion of a thermal state discussed in Section~\ref{sec: class_instability}. 
The stationarity condition~\eqref{eq: deri_phase_func_t_DoS_IHO} then demands $\cosh (\upomega \tau_{0} / 2) = 1$. 
Employing the identity $\cosh (t + \ri s) = \cosh t \cos s + \ri \sinh t \sin s$ and the fact that that both $h_{-}$ and $E$ are real, we obtain 
\begin{equation}
    t_{0} = 0, \quad s_{0} = kT, \quad k \in \Z,     
\end{equation}
i.e., the critical point $\tau_{0}$ is purely imaginary $\tau_{0} = \ri kT$.  
The stationarity conditions~\eqref{eq: deri_phase_func_x_DoS_IHO} and~\eqref{eq: deri_phase_func_xi_DoS_IHO} are analogous to those (\eqref{eq: deri_phase_func_x_DoS_SHO} and~\eqref{eq: deri_phase_func_xi_DoS_SHO}) in the case of a SHO. 
Therefore, $\phi_{\tau}^{-}$ satisfies the Hamilton-Jacobi equation at $\gamma_{l}^{-} (\ri kT) \in \Sigma_{E}^{-}$; see~\eqref{eq: const_energy_periodic_geodesic_SHO} and~\eqref{eq: cst_energy_surface_IHO}. 

A straightforward computation yields $\det \Hess_{\gamma_{l}^{-} (\ri kT)} \phi_{\tau}^{-} = 2 \partial_{x} h_{-} \big( \gamma_{l}^{-} (\ri kT) \big) \partial_{\xi} h_{-} \big( \gamma_{l}^{-} (\ri kT) \big)$ $\neq 0$. 
That is, $\phi_{\tau}^{+} + \tau E$ is a non-degenerate phase function and the leading term in the stationary phase approximation yields 
\begin{equation} \label{eq: DoS_IHO_WKB}
    \uprho_{-} (E) 
    \propto  
    \frac{1}{2 \pi \hbar \upomega} \sum_{l=1, 2} \sum_{\gamma_{l}^{-} (kT)} \re^{- kTE / \hbar} \re^{\ri \pi \km_{l}^{-} / 4}, 
\end{equation}
where the constant of proportionality is a dimensionless constant and the signature of $\Hess_{\gamma_{l}^{-} (\ri kT)} (\phi_{\tau}^{-} + \tau E)$ is the Maslov index $\km_{l}^{-}$ of $\gamma_{l}^{-}$.  

%
%
%
%
%
%
%
%
%
%
\section{Conclusion} 

The primary result (Section~\ref{sec: DoS_IHO}) of this article is the computation of the semi-classical density of states $\uprho_{-} (E)$ of an IHO on a circle $\bbS$ without employing analytic continuation and Gutzwiller trace formula. 
The first exponential term in~\eqref{eq: DoS_IHO_WKB} is 
precisely\footnote{The $2 \pi$ factor 
in~\cite[(15)]{Dalui_PRD_2020},\cite[(C5)]{Dalui_ii_PRD_2020} 
is missing in~\eqref{eq: DoS_IHO_WKB} because of the different convention used for the Fourier transform.
} 
that 
in~\cite[(15)]{Dalui_PRD_2020} 
(see 
also~\cite[(C5)]{Dalui_ii_PRD_2020})  
but there is an additional summation $\sum_{l}$ in our expression~\eqref{eq: DoS_IHO_WKB} as well as the Maslov index $\km_{l}^{-}$ differs. 
The additional summation over $l = 1, 2$ in~\eqref{eq: DoS_IHO_WKB} accounts the fact that, for each energy $E > 0$, there are two distinct phase-space trajectories $\gamma_{l}^{-}$ of an IHO. 
This summation is missing in the preceding references because a SHO has only one trajectory $\gamma^{+}$ for each $E > 0$. 
Moreover, the Maslov index $\km^{+}$ for a SHO differs from that $\km_{l}^{-}$ of an IHO since $\gamma^{+}$ is not topologically equivalent (homeomorphism) to $\gamma_{l}^{-}$. 

At this point, let us emphasise that it is essential to consider $\bbS \times \R$ as the phase-space and $L^{2} (\bbS)$ as the Hilbert space of an IHO in contrast to $\R \times \R$ and $L^{2} (\R)$ for those of a SHO. 
This is because an IHO Hamiltonain $\hat{H}_{-}$ on $L^{2} (\R)$ has absolutely continuous real spectrum~\eqref{eq: spec_IHO} --- a fact that cannot be obtained by analytically continuing the spectrum~\eqref{eq: EV_SHO} of a SHO because that yields complex eigenvalues~\eqref{eq: EV_IHO} for $\hat{H}_{-}$. 
As a consequence, one must consider $\hat{H}_{-}$ as an operator on $L^{2} (\bbS)$ instead of $L^{2} (\R)$ in order to ensure discrete eigenvalues required to make sense the notion of density of states. 
Furthermore, time must be considered as a complex parameter $\tau$ (in the sense of~\eqref{eq: strip_holo}) for an IHO in order to guarantee closed trajectories $\gamma_{l}^{-} (\tau)$ on the phase-space $\bbS \times \R$ as well as defining the notion of a thermal state. 
Physically, this accounts to trade-off between the repulsive nature of an IHO trajectory and periodicity. 

To elucidate the underlying reasons behind the above-mentioned physical and mathematical gaps in the work of Dalui and his 
collaborators~\cite{Dalui_PRD_2020, Dalui_ii_PRD_2020}, 
we have computed the density of states $\uprho_{+} (E)$ of a SHO employing the stationary phase approximation in Section~\ref{sec: DoS_SHO}. 
It is now evident from~\eqref{eq: DoS_SHO_WKB} and~\eqref{eq: DoS_IHO_WKB} that the urge to infer~\eqref{eq: DoS_IHO_WKB} from~\eqref{eq: DoS_SHO_WKB} via the analytic continuation $\upomega \mapsto \ri \upomega$ is not completely unnatural. 
A naive attempt (as in the preceding references) will result in an incorrect Maslov factor and miss the summation over $l$ in~\eqref{eq: DoS_IHO_WKB} but these can be avoided by a careful examination of the phase-space dynamics under the analytic continuation. 
It is the spectral theory (Section~\ref{sec: spec_th}) of $\hat{H}_{-}$ where the analytic continuation-based approach fails completely and yields physically unreasonable results like complex eigenvalues~\eqref{eq: EV_IHO} of $\hat{H}_{-}$. 

We close this section by summarising the preceding discussion as follows. 
Our expression~\eqref{eq: DoS_IHO_WKB} of the semi-classical density of states of an IHO reproduces the primary result by Dalui and his 
collaborators~\cite{Dalui_PRD_2020, Dalui_ii_PRD_2020}, 
points out and corrects their mistakes in the Maslov index and energy-degeneracy of periodic orbits $\gamma_{l}^{-} (\tau)$, and clarifies the gaps in the spectral properties of an IHO Hamiltonian $\hat{H}_{-}$ used in their article. 
In the future, we would like to propound this picture in a quantum field theoretic setting. 
At this point, it is not evident whether the stationary phase lemma is adequate for a field-theoretic generalisation. 
Also, the time-evolution operator of the Klein-Gordon equation, for instance, is quite different from an IHO Mehler kernel~\eqref{eq: Mehler_kernel_SIHO} used in our present investigation.  
We plan to address all these in the forthcoming article. 

%
%
%
%
%
%
%
%
%
%
\section*{Acknowledgement} 

A.C. would like to thank Dmitry Jakobson for supervising his undergraduate summer project on quantum chaos, and for numerous insightful discussions on spectral theory and semi-classical analysis. 
O.I. is financially supported by Deutsche Forschungsgemeinschaft (DFG) --- Project Number 546644682: „Feynman-Green-Operatoren f\"{u}r Dirac-Operatoren mit nichtlokalen Randbedingungen“. 

%
%
%
%
%
%
%
%
%
%
\appendix
\counterwithin*{equation}{section}
\renewcommand\theequation{\thesection.\arabic{equation}}
\section{Gel$'$fand triplets}
\label{sec: Gelfand_triplet}

In this appendix, we present the necessary background on functional analysis required in this article. 
Mathematical details are available, for instance, in the 
treatises~\cite{Gelfand_AMS_1964, Reed_I} 
whereas we refer to the 
textbook~\cite{Bohm_Springer_2001} 
for quantum mechanical aspects. 
To begin with, let $\kD$ be a topological vector space over the field $\C$.  
We denote the space of continuous $\C$-linear (resp. anti-linear) functionals $u$ (resp. $\bar{u}$) $: \kD \to \C$ by $\kD'$ (resp. $\bar{\kD}'$). 
Recall that $\bar{\kD}' \equiv \overline{\kD'} := \{ \bar{u} ~|~ u \in \kD' \}$ is the conjugate vector space of $\kD'$ and there is a canonical \textit{anti}-linear isomorphism $\kD' \ni u \mapsto \bar{u} \in \bar{\kD}'$, where $\bar{u} (\phi) := \overline{u (\phi)}$ for any $u \in \kD'$ and $\phi \in \kD$.  
Let $L : \kD \to \kD$ be a continuous linear operator on $\kD$.  
Then its \textit{formal} adjoint $L' : \bar{\kD}' \to \bar{\kD}'$ is defined by 
\begin{equation} \label{eq: def_formal_adjoint_op}
    (L' u) (\phi) := u (L \phi) 
\end{equation}
for any $u \in \bar{\kD}'$ and $\phi \in \kD$. 
The operator $L$ is called \textit{formally} self-adjoint if $L = L'$. 
Recall that a distribution $u \in \bar{\kD}'$ is called a \textit{generalised} eigenvector of $L$ with eigenvalue $\lambda$ if $L u = \lambda u$, i.e., $u (L' \phi) = u (\bar{\lambda} \phi) = \lambda u (\phi)$ for any $\phi \in \kD$. 
Note that the eigenvalues of a formally self-adjoint operator are \textit{not} necessarily real. 

Let $(\cH, \scalarProdTwo{\cdot}{\cdot})$ be a (complex separable) Hilbert space where the inner product $\scalarProdTwo{\cdot}{\cdot}$ is anti-linear (resp. linear) in the first (resp. second) argument.  
By an operator $L$ on $\cH$, one means the linear map $L : \kD (L) \to \cH$ whose domain $\kD (L)$ is a dense subset of $\cH$. 
If there exists an element $\psi \in \kD (L) \setminus \{ 0 \}$ such that $L \psi = \lambda \psi$ for some $\lambda \in \C$, then $\psi$ is called an eigenvector with eigenvalue $\lambda$. 
The set $\ptSpec L$ of all eigenvalues is called the point spectrum whereas the set of complex numbers $\lambda$ for which $L - \lambda \one$ does not admit a bounded inverse is called the spectrum $\Spec L$. 
Obviously $\ptSpec L \subseteq \Spec L$. 
Let $\kD (L^{*})$ be the set of elements $\phi \in \cH$ such that there exists some $\tilde{\phi} \in \cH$ satisfying $\langle \tilde{\phi} | \psi \rangle = \scalarProdTwo{\phi}{L \psi}$ for any $\psi \in \kD (L)$. 
Then the adjoint $L^{*}$ of $L$ is defined by $L^{*} \phi = \tilde{\phi}$. 
Recall that $L$ is called symmetric if 
\begin{equation} \label{eq: def_symm_op}
    \scalarProdTwo{L \psi}{\phi} = \scalarProdTwo{\phi}{L \psi} 
\end{equation}
for all $\psi, \phi \in \kD (L)$, self-adjoint if $L = L^{*}$, and essentially self-adjoint if its closure is self-adjoint.  
A well-known fact about self-adjoint operators is that their spectrum must be \textit{real}. 

We now consider a topological vector space $(\kD, \scalarProdTwo{\cdot}{\cdot})$ endowed with an inner product $\scalarProdTwo{\cdot}{\cdot}$. 
Let $\cH$ be the completion of $\kD$. 
This yields the Hilbert space $(\cH, \scalarProdTwo{\cdot}{\cdot})$ whose topological dual is denoted by $\cH'$.  
One identifies $\cH$ with $\cH'$ owing to the canonical \textit{anti}-linear isometry given by the (Fr\'{e}chet-)Riesz lemma and then the triplet of spaces $\kD \subset \cH \equiv \cH' \subset \bar{\cD}'$ is called a Gel$'$fand triplet (also known as a Rigged Hilbert space). 
For instance, the continuous inclusions 
\begin{equation} \label{eq: Schwartz_subset_LTwo_subset_tempared}
    \kD := \cS (\R) \hookrightarrow \cH := L^{2} (\R) \hookrightarrow \kD' := \cS' (\R) 
\end{equation}
is an example of Gel$'$fand triplet, where $\cS (\R), L^{2} (\R)$, and $\cS' (\R)$ are the spaces of Schwartz functions, square-integrable functions, and tempered distributions, respectively. 
Elements $\psi \in \kD$ (resp. $\cH$) and $u \in \kD'$ (resp. $\cH'$) are called kets and bras, respectively. 
In Dirac bra-ket notation, $\psi = | \psi \rangle$ and $u = \langle \psi |$. 
By the (Fr\'{e}chet-)Riesz lemma, for each ket $\psi \in \cH$, there exists a unique bra $u \in \cH'$, and vice-versa. 
However, \textit{not} every bra in $\bar{\kD}'$ admits a ket in either $\kD$ or $\cH$. 

One often needs to deal with unbounded self-adjoint operators $L$ on $\cH$ in quantum mechanics. 
Recall that an operator $L$ is bounded if of $\sup_{\psi \in \kD (L) \setminus \{ 0 \}} \| L \psi \| / \| \psi \| < \infty$, and unbounded, otherwise. 
The latter class of operators can only be defined on a dense subset of $\cH$. 
In the Gel$'$fand triplet formalism, one chooses $\kD$ in such a way that it is contained in the domain of all relevant operators and their compositions, and is invariant under the action of those operators. 
To elucidate the functional analytical aspects, let us consider the multiplication operator with $x$   
\begin{equation}
    \hat{x} : \kD (\hat{x}) := \{ \psi \in \cH ~|~x \psi \in \cH \} \to \cH, \quad \psi \mapsto 
    (\hat{x} \psi) (x) := x \psi (x)    
\end{equation}
on $\cH := L^{2} (\R)$. 
It can be shown that $\hat{x}$ is unbounded and self-adjoint. 
Observe that $x \psi$ even makes sense when $\psi$ is a distribution. 
For instance, consider the multiplication operator $\hat{x}_{\upharpoonright \cS (\R)} : \cS (\R) \to \cS (\R)$ restricted to $\cS (\R) \subset \kD (\hat{x})$. 
The position operator is then formally self-adjoint and it acts on $\cS' (\R)$ by duality. 
If $\delta_{a} \in \bar{\cS}'$ be the Dirac delta distribution concentrated at $x = a$ then 
$(\hat{x} \delta_{a}) (\phi) = \delta_{a} (\hat{x} \phi) = \delta_{a} (x \phi) = (x \delta_{a}) (\phi) = a \delta_{a} (\phi)$ for any $\phi \in \cS (\R)$. 
This entails that $\delta_{a}$ is a generalised eigenvector of $\hat{x}$ since $\delta_{a} \notin \cH$. 

%
%
%
%
%
%
%
%
%
%
\section{Pseudodifferential operators}
\label{sec: PsiDO}

Essentials of semi-classical pseudodifferential operators pertinent to this article have been recalled in this appendix 
following~\cite[Chap. 7]{Dimassi_CUP_1999},~\cite[Chap. 4]{Zworski_AMS_2012},~\cite[Chap. 1]{Taylor_AMS_1986}.   
Let $d \in \N$ and $\alpha = (\alpha_{1}, \ldots, \alpha_{d}) \in \N_{0}^{d}$ where $ \N_{0} := \N \cup \{0\}$. 
A partial derivative in the multi-index notion is given by $D_{\xi}^{\alpha} := (- \ri)^{|\alpha|} \, \nicefrac{\partial^{|\alpha|}}{ \partial \xi_{1}^{\alpha_{1}} \ldots \partial \xi_{n}^{\alpha_{d}} }$, where $|\alpha| := |\alpha_{1}| + \ldots + |\alpha_{n}|$. 
We denote the space of complex-valued smooth functions on the cotangent bundle $\coTanRd = \Rd \times \Rd$ of $\Rd$ by $C^{\infty} (\Rd \times \Rd)$. 

Let $(0, \boldsymbol{h}] \ni h \mapsto \fp (h; \cdot) \in C^{\infty} (\Rd \times \Rd)$ be a function for some $\boldsymbol{h} > 0$. 
The space $S_{1}^{m} (\Rd \times \Rd)$ of semi-classical Shubin symbols (also known as the semi-classical symbols of harmonic oscillator type) on $\Rd \times \Rd$ of order (at most) $m \in \R$ is defined as the set of all functions $\fp$ on $(0, \boldsymbol{h}] \times \Rd \times \Rd$ such that the estimation holds 
\begin{equation*}
    |D_{x}^{\beta} D_{\xi}^{\alpha} \fp (h; x, \xi)| \leq c_{\alpha \beta} (1 + |x| + |\xi|)^{m -|\alpha| - |\beta|} 
\end{equation*}
uniformly in $h$ for some constant $c_{\alpha \beta}$ depending on $\alpha, \beta \in \N_{0}^{d}$. 
Then the classical symbol $S^{m} (\Rd \times \Rd)$ on $\Rd \times \Rd$ of order $m$ is defined as the set of all $\fp \in S_{1}^{m} (\Rd \times \Rd)$ such that 
\begin{equation} \label{eq: asymptotic_expansion}
    \fp (h; x, \xi) - \sum_{k=0}^{N-1} h^{k} p_{k} \xxi \in h^{N} S_{1}^{m - 2N} (\Rd \times \Rd)
\end{equation}
for each $N \in \N$. 
Here, for each $k$, $p_{k} \xxi$ is smooth and homogeneous in $\xxi$ for $|x|^{2} + |\xi|^{2} \geq 1$ of degree $m - 2k$. 
Note that the summation in~\eqref{eq: asymptotic_expansion} usually does not converge, and it is called asymptotic summation, symbolised as $\fp \sim \sum_{k} h^{k} p_{k}$. 

\begin{subequations}
    A semi-classical pseudodifferential operator $P \in \PsiDO{m}$ on $\Rd$ of order $m$ is, per se, a continuous linear operator  
    \begin{equation}
        P : \cS (\Rd) \to \cS (\Rd), \quad u \mapsto (Pu) (x) = \fP (h; x, y) \big( u (y) \big) 
    \end{equation}
    whose Schwartz kernel in the Weyl calculus is of the form 
    \begin{equation} \label{eq: def_kernel_PsiDO}
        \fP (h; x, y) := \int_{\Rd} \fp \left( h; \frac{x+y}{2}, \xi \right) \re^{\nicefrac{\ri}{h} \, (x-y) \xi} \frac{\rd \xi}{(2 \pi h)^{d}}. 
    \end{equation}
    Note that the preceding integral is an oscillatory integral and hence it is to be understood as a distribution.  
\end{subequations}
By duality~\eqref{eq: def_formal_adjoint_op}, $P$ acts on Schwartz distributions: $P' : \cS' (\Rd) \to \cS' (\Rd)$. 

The function $\fp$ in~\eqref{eq: def_kernel_PsiDO} is called the (total) Weyl symbol of $P$, and it does not have an invariant meaning. 
In contrast, the leading-order contribution in the asymptotic summation has an intrinsic meaning, and it is called the principal symbol 
\begin{equation}
    \symb{P} : \PsiDO{m} \to S^{m} (\Rd \times \Rd), \quad P \mapsto \symb{P} \xxi := p_{0} \xxi  
\end{equation}
of $P$. 
Roughly speaking, principal symbols correspond to classical observables whilst the corresponding zeroth-order pseudodifferential operators correspond to quantum observables. 
To be precise, the Weyl quantisation is the map
\begin{equation} \label{eq: Weyl_quantisation}
    C^{\infty} (\Rd \times \Rd, \R) \ni p \mapsto P \in \PsiDO{0} \quad | \quad \symb{P} = p.   
\end{equation}



\bibliography{ref}

@article{Barton_AnnPhys_1986,
 title=		{Quantum mechanics of the inverted oscillator potential},
 journal=	{Ann. Phys.},
 volume=	{166},
 number=	{2},
 pages=		{322--363},
 year=		{1986},
 issn=		{0003-4916},
 doi=		{https://doi.org/10.1016/0003-4916(86)90142-9},
 url=		{https://www.sciencedirect.com/science/article/pii/0003491686901429},
 author=	{Barton, G.}
}

@BOOK{Bohm_Springer_2001,
   author=	{Bohm, Arno},
   title=	{Quantum Mechanics: Foundations and Applications},
   publisher=	{Springer-Verlag},
   edition=	{Third},
   year=	{2001},
   address=	{India}
}

@book{Bratteli_Springer_1997,
 title=    {Operator Algebras and Quantum Statistical Mechanics - II},
 author=   {Bratteli, Ola and Robinson, Derek William},
 volume=   {},
 edition=  {Second},
 isbn=     {978-3-540-61443-2},
 lccn=     {},
 series=   {Theoretical and Mathematical Physics},
 url=      {http://www.springer.com/us/book/9783540614432},
 year=     {1997},
 address=  {Berlin Heidelberg},
 publisher={Springer-Verlag}
}

@article{Dalui_PRD_2020,
  title = {Horizon induces instability locally and creates quantum thermality},
  author = {Dalui, Surojit and Majhi, Bibhas Ranjan and Mishra, Pankaj},
  journal = {Phys. Rev. D},
  volume = {102},
  issue = {4},
  pages = {044006},
  numpages = {6},
  year = {2020},
  month = {Aug},
  publisher = {American Physical Society},
  doi = {10.1103/PhysRevD.102.044006},
  url = {https://link.aps.org/doi/10.1103/PhysRevD.102.044006}, 
  eprint={1910.07989},
  archivePrefix={arXiv},
  primaryClass={gr-qc}
}

@article{Dalui_ii_PRD_2020,
  title = {Near-horizon local instability and quantum thermality},
  author = {Dalui, Surojit and Majhi, Bibhas Ranjan},
  journal = {Phys. Rev. D},
  volume = {102},
  issue = {12},
  pages = {124047},
  numpages = {19},
  year = {2020},
  month = {Dec},
  publisher = {American Physical Society},
  doi = {10.1103/PhysRevD.102.124047},
  url = {https://link.aps.org/doi/10.1103/PhysRevD.102.124047}, 
  eprint={2007.14312},
  archivePrefix={arXiv},
  primaryClass={gr-qc}
}

@incollection{Derezinski_JEDP_1993,
 author=	{Derezinski, Jan},
 title=		{Some remarks on {Weyl} pseudodifferential operators},
 series=	{Days \partial differential equations},
 eid=		{12},
 pages=		{1--14},
 publisher=	{Ecole polytechnique},
 year=		{1993},
 doi=		{10.5802/jedp.449},
 url=		{https://www.numdam.org/articles/10.5802/jedp.449/}
}

@book{Dimassi_CUP_1999, 
 place= 	{Cambridge}, 
 series=	{London Mathematical Society Lecture Note Series}, 
 title=		{Spectral Asymptotics in the Semi-Classical Limit}, 
 publisher=	{Cambridge University Press}, 
 author=	{Dimassi, M. and Sjostrand, J.}, 
 year=		{1999}, 
 doi=		{https://doi.org/10.1017/CBO9780511662195}
}

@Article{Duistermaat_InventMath_1975,
 author=	{Duistermaat, J. J. and Guillemin, V. W.},
 title=		{The spectrum of positive elliptic operators and periodic bicharacteristics},
 journal=	{Invent. math.},
 year=		{1975},
 month=		{Feb},
 day=		{01},
 volume=	{29},
 number=	{1},
 pages=		{39--79},
 issn=		{1432-1297},
 doi=		{10.1007/BF01405172},
 url=		{https://doi.org/10.1007/BF01405172}
}

@article{Fulling_PR_1987,
 title=		{Temperature, periodicity and horizons},
 journal=	{Phys. Rep.},
 volume=	{152},
 number=	{3},
 pages=		{135--176},
 year=		{1987},
 issn=		{0370-1573},
 doi=		{https://doi.org/10.1016/0370-1573(87)90136-0},
 url=		{https://www.sciencedirect.com/science/article/pii/0370157387901360},
 author=	{Fulling, S.A. Fulling and Ruijsenaars, S.N.M.}
}

@article{Gutzwiller_JMP_1971,
 author=	{Gutzwiller, Martin C.},
 title=		{Periodic Orbits and Classical Quantization Conditions},
 journal=	{J. Math. Phys.},
 volume=	{12}, 
 number=	{3},
 pages=		{343-358},
 year=		{1971},
 doi=		{10.1063/1.1665596},
 URL=		{https://doi.org/10.1063/1.1665596}
}

@BOOK{Gutzwiller_Springer_1990,
 author=	{Gutzwiller, Martin C.},
 title=		{Chaos in Classical and Quantum Mechanics},
 series=	{Interdisciplinary Applied Mathematics},
 volume=	{1},
 publisher=	{Springer-Verlag},
 address=	{New York},
 year=		{1990},
 doi=		{10.1007/978-1-4612-0983-6},
 url=		{https://www.springer.com/gb/book/9780387971735}
}

@ARTICLE{Gutzwiller_Scholarpedia_2007,
 AUTHOR=	{Gutzwiller, Martin},
 TITLE=		{Quantum chaos},
 YEAR=		{2007},
 JOURNAL=	{Scholarpedia},
 VOLUME=	{2},
 NUMBER=	{12},
 PAGES=		{3146},
 DOI=		{doi:10.4249/scholarpedia.3146}
}

@article{Hawking_CMP_1975,
 author=	{Hawking, S. W.},
 fjournal=	{Communications in Mathematical Physics},
 journal=	{Commun. Math. Phys.},
 number=	{3},
 pages=		{199--220},
 publisher=	{Springer},
 title=		{Particle creation by black holes},
 url=		{http://projecteuclid.org/euclid.cmp/1103899181},
 volume=	{43},
 year=		{1975}
}

@article{Hollands_PR_2015,
 title=	     	{Quantum fields in curved spacetime},
 journal=    	{Phys. Rep.},
 volume=     	{574},
 number=     	{},
 pages=	     	{1-35},
 year=	     	{2015},
 note=	     	{},
 issn=	     	{0370-1573},
 doi=	     	{http://dx.doi.org/10.1016/j.physrep.2015.02.001},
 url=	     	{http://www.sciencedirect.com/science/article/pii/S0370157315001416},
 author=     	{Hollands, Stefan and Wald, Robert M.},
 archivePrefix= {arXiv},
 eprint       = {1401.2026},
 primaryClass = {gr-qc}
}

@article{Hoermander_MathZ_1995,
 title=		{Symplectic classification of quadratic forms, and general Mehler formulas},
 journal=	{Math. Z.},
 volume=	{219},
 pages=		{413--449},
 year=		{1995},
 doi=		{https://doi.org/10.1007/BF02572374},
 url=		{https://doi.org/10.1007/BF02572374},
 author=	{H\"{o}rmander, Lars},
}

@article{Hashimoto_PRD_2017,
  title = {Universality in chaos of particle motion near black hole horizon},
  author = {Hashimoto, Koji and Tanahashi, Norihiro},
  journal = {Phys. Rev. D},
  volume = {95},
  issue = {2},
  pages = {024007},
  numpages = {13},
  year = {2017},
  month = {Jan},
  publisher = {American Physical Society},
  doi = {10.1103/PhysRevD.95.024007},
  url = {https://link.aps.org/doi/10.1103/PhysRevD.95.024007}, 
  archivePrefix=	{arXiv},
 eprint       =	{1610.06070},
 primaryClass =	{hep-th}
}

@incollection{Kay_EncyMathPhys_2025,
 title={Quantum Field Theory in Curved Spacetime},
 editor={Szabo, Richard and Bojowald, Martin},
 booktitle={Encyclopedia of Mathematical Physics},
 publisher={Academic Press},
 edition={Second Edition},
 address={Oxford},
 pages={357-381},
 year={2025},
 isbn={978-0-323-95706-9},
 doi={https://doi.org/10.1016/B978-0-323-95703-8.00085-9},
 url={https://www.sciencedirect.com/science/article/pii/B9780323957038000859},
 author={Kay, Bernard S.}, 
 eprint={2308.14517},
 archivePrefix={arXiv},
 primaryClass={gr-qc},
}

@article{Meinrenken_ReptMathPhys_1992,
 title=		{Semiclassical principal symbols and {G}utzwiller's trace formula},
 journal=	{Rep. Math. Phys.},
 volume=	{31},
 number=	{3},
 pages=	    	{279--295},
 year=		{1992},
 issn=		{0034-4877},
 doi=		{https://doi.org/10.1016/0034-4877(92)90019-W},
 url=		{http://www.sciencedirect.com/science/article/pii/003448779290019W},
 author=	{Meinrenken, Eckhard}
}

@book{Reed_I,
 author=	{Reed, Michael and Simon, Barry},
 title=		{Functional Analysis},
 publisher=	{Acadmic Press},
 address=	{USA},
 series=	{Methods of Modern Mathematical Physics},
 volume= 	{I},
 year=		{1980}
}

@book{Gelfand_AMS_1964,
 author=	   {Gel'fand, I. M. and Vilenkin, N. Ya.},
 title=		{Generalized Functions, Volume 4: Applications of Harmonic Analysis},
 edition=	{AMS Reprint: 2016},
 publisher=	{AMS Chelsea Publishing},
 address=	{USA},
 year=		{1964},
 url=		{https://bookstore.ams.org/chel-380-h}
}

@book{Taylor_AMS_1986,
 author=	{Taylor, Michael E.},
 title=		{Noncommutative Harmonic Analysis},
 series=	{Mathematical Surveys and Monographs},
 volume=	{22},
 publisher=	{American Mathematical Society},
 address=	{USA},
 year=		{1986},
 doi=		{https://doi.org/10.1090/surv/022}
}

@book{Rudolph_Springer_2013,
 title=		{Differential Geometry and Mathematical Physics - {I}: Manifolds, Lie Groups and Hamiltonian Systems},
 author=	{Rudolph, Gerd and Schmidt, Matthias},
 volume=	{},
 doi=		{10.1007/978-94-007-5345-7},
 series=	{Theoretical and Mathematical Physics},
 url=		{https://www.springer.com/gb/book/9789400753440},
 year=		{2013},
 address=	{Netherlands}, 
 publisher=	{Springer}
}

@CONFERENCE{Uribe_Cuernavaca_1998,
 AUTHOR=	{Uribe, Alejandro}, 
 Editor=	{P\'{e}rez-Esteva, Salvador and Villegas-Blas, Carlos},
 TITLE=	    	{Trace Formulae},
 BOOKTITLE=	{First Summer School in Analysis and Mathematical Physics: Quantization, the {S}egal-{B}argmann Transform and Semiclassical Analysis},
 YEAR=		{2000},
 Pages=	    	{61--90},
 Publisher=	{American Mathematical Society},
 Address=	{Cuernavaca Morelos, Mexico},
 Month=	    	{1998},
 url=		{http://www.ams.org/books/conm/260/},
 doi=		{10.1090/conm/260},
 note= 	{Contemporary Mathematics \textbf{260} (2000)}
}

@Article{Wald_LivRevRel_2001,
 author=	{Wald, Robert M.},
 title=		{The Thermodynamics of Black Holes. Living Rev. in Rel. 4, 6 (2001). },
 journal=	{Liv. Rev. Relativity},
 year=		{2001},
 volume=	{4},
 pages=		{6},
 doi=		{http://dx.doi.org/10.12942/lrr-2011-8},
 url=		{http://dx.doi.org/10.12942/lrr-2011-8},
 archivePrefix= {arXiv},
 eprint= 	{9912119},
 primaryClass= 	{gr-qc}
}

@book{Weidmann_Springer_2003,
 title=	{Lineare Operatoren in Hilberträumen Teil II: Anwendungen},
 author=	{Weidmann, Joachim},
 series=	{Mathematische Leitf\"{a}den},
 doi={https://doi.org/10.1007/978-3-322-80095-4},
 url=		{https://link.springer.com/book/10.1007/978-3-322-80095-4},
 year=		{2003},
 publisher=	{Vieweg+Teubner Verlag}, 
 address=	{Wiesbaden}
}

@book{Zworski_AMS_2012,
  title=	{Semiclassical Analysis},
  author=	{Zworski, Maciej},
  series=	{Graduate Studies in Mathematics},
  volumn=	{138}, 
  url=		{https://bookstore.ams.org/gsm-138},
  year=		{2012},
  publisher=	{American Mathematical Society}, 
  address=	{USA}
}
\end{document}